\documentclass[showkeys,prb,twocolumn]{revtex4-1}
\usepackage{amsmath}
\usepackage{color}
\usepackage{hyperref}
\usepackage{graphicx}
\usepackage{setspace}

\newcommand{\VEC}[1]{{\boldsymbol{ #1}}}

\newcommand{\etal}{{\it et al.}}

\newcommand{\Gru}{{Gr\"uneisen}}
\newcommand{\GP}{{Gr\"uneisen parameter}}
\newcommand{\Fig}{{Fig.}}
\newcommand{\Eq}{{Eq.}}
\newcommand{\Eqs}{{Eqs.}}
\newcommand{\Table}{{Table}}
\newcommand{\Sec}{{Sec.}}
\newcommand{\nbs}{{NbS$_3$}}
\newcommand{\nbsfour}{{NbS$_3$-IV}}
\newcommand{\invcm}{{cm$^{-1}$}}

\newcommand{\be}{\begin{equation}} \newcommand{\ee}{\end{equation}}
\newcommand{\bea}{\begin{eqnarray}} \newcommand{\eea}{\end{eqnarray}}
\newcommand{\td}[2] {{ \frac{d #1}{d #2} }}
\newcommand{\half}{{\frac{1}{2}}}
\newcommand{\kBol}{ k_{{\scriptscriptstyle B}}}
\newcommand{\lp}{lattice-parameter}

\newcommand{\azero}{a}
\newcommand{\bzero}{b}
\newcommand{\czero}{c}
\newcommand{\betazero}{\beta}

\begin{document}
\title{
Large thermal anisotropy in monoclinic niobium trisulfide: A thermal expansion tensor study
}
\author{Chee Kwan Gan}
\email{ganck@ihpc.a-star.edu.sg}
\affiliation{Institute of High Performance Computing, 1 Fusionopolis Way, \#16-16 Connexis, Singapore 138632}
\author{Kun Ting Eddie Chua}
\affiliation{Institute of High Performance Computing, 1 Fusionopolis Way, \#16-16 Connexis, Singapore 138632}
\date{4 Jan 2019} 

\begin{abstract}
We present a method based on the \Gru{} formalism to
calculate the thermal expansion coefficient (TEC) tensor that is
applicable to any crystal system, where the number of phonon
calculations associated with different deformations scales linearly
with the number of lattice parameters.  
Compared to simple high-symmetry systems such 
as cubic or hexagonal systems, a proper consideration of
low-symmetry systems such as monoclinic or triclinic crystals demands
a clear distinction between the TEC tensor and the \lp{} TECs along
the crystallographic directions.  The latter is more complicated and 
it involves integrating the
equations of motion for the primitive lattice vectors, with input from
the TEC tensor.  
A first-principles study of the TEC is carried out for the first time on a 
monoclinc crystal, where
we unveil high TEC anisotropies in a
recently reported monoclinic phase of niobium trisulfide (\nbs{})
crystal with a relatively large primitive cell (32 atoms per cell) using density-functional theory.
We find the occurrence of a negative TEC tensor component 
is largely due to the mechanical property rather than the anharmonic effect, contrary to the common belief.
Our theoretical treatment of the monoclinic system with a single
off-diagonal tensor element could be routinely generalized to
any crystal system, including the lowest-symmetry triclinic system with three
off-diagonal tensor elements.
\end{abstract}

\maketitle

\section{Introduction}

Thermal expansion of materials is a direct manifestation of the
anharmonic effects inherent in a solid. A good understanding of the
phonon dynamics affected by temperature-induced crystal deformations
is often required for improved device
applications\cite{Goodwin08v319,Lin11v83,Kim12v100}.  
So far {\it ab initio} studies of anharmonic effects in high-symmetry
crystals such as cubic and hexagonal systems have been carried out
via the quasiharmonic approximation (QHA)
\cite{Mounet05v71,Stoffel15v27}.  The
application of QHA on low-symmetry crystals is extremely challenging because
many sets of time-consuming phonon calculations are required to search
for the free energy minimum in a much larger lattice-parameter
space, which may explain why, to the best of our knowledge, no studies 
have been carried
out on, say, a monoclinic system. Even if a direct free energy minimization via QHA could be
carried out, it may be still difficult to understand the underlying
physics without interpreting the phonon frequency shifts due to a
crystal deformation, phonon density of states, mechanical properties
such as elastic constants and elastic compliances, and heat
capacities.  Since many technologically
important materials are also found in low
crystal symmetries, e.g., hybrid
perovskites\cite{Calabrese91v113,Du17v56,Giovanni18v5} and
ceramics\cite{Zhao02v65,Haggerty14v97}, it is desirable to be
able to address the anharmonicity in these materials through other
means.  Recently, the \Gru{} approach has between applied to 
high-symmetry crystal systems and
good agreements between theoretical
predictions and experimental results
\cite{Ding15v5,Liu17v121,Gan18v151}.
In this work, we extend the utility of the \Gru{}
formalism\cite{Gruneisen26v10,Schelling03v68} to handle low-symmetry
crystals.  As will be shown later in this paper, dealing with
low-symmetry crystals requires a clear distinction to be made between
the thermal expansion coefficient (TEC) tensor and the
lattice-parameter TECs. These two concepts, unfortunately, coincide
with each other in high-symmetry crystals such as the cubic crystal.
Other methods to calculate TECs
are the vibrational self-consistent-field method\cite{Monserrat13v87}
and the self-consistent quasiharmonic
approximation\cite{Huang16v120}.  We should note that the knowledge of
the third-order interatomic force constants\cite{Barron74,Shiomi11v84}
can be used directly to calculate the \GP{}s for the evaluation of
TECs.  Another recent work \cite{Lee17v96} outlines a reverse process
where numerically computed \GP{}s could be used to extract the
third-order interatomic force constants for the calculations of
thermal conductivity.

The monoclinic \nbs{}-IV\cite{Bloodgood18v6} (with four lattice parameters)
belongs to a class of semiconducting transition metal chalcogenides.
Transition metal chalcogenides have received attention due to their
potential to replace traditional semiconductor technologies. They have
been investigated for their electronic and optical
properties, and their potential use in electronic
devices\cite{Ribeiro18v97}.  The extensive interest in metal
chalcogenides\cite{Zhao11v84,Chong14v90} arises from the low-dimensionality of the structure.  As
a prototypical example of metal trichalcogenides (MX$_3$), niobium
trisulfide (\nbs) comprises of [NbS$_6$] units, where the sulfur
atoms are arranged in a triangular prism around a central Nb
atom\cite{vanSmaalen88v38}.  Neighboring trigonal prisms share a
triangle face, thus fusing individual units to form an infinite chain
along the $b$ axis.  The chains are held together by additional
covalent bonds to form bilayer sheets which interact via van der Waals
forces.  Experimentally, \nbs{} has been found to crystallize in six
polymorphs: I, II, III, IV, V and a high-pressure phase.  The first
polymorph of \nbs, \nbs-I, first identified using single crystal x-ray
diffraction\cite{Jellinek60v185,Rijnsdorp78v25}, belongs to space
group $P\bar{1}$.  Using electron diffraction studies, \nbs-II was the
second polymorph to be identified, although the structure remains
unknown to date.  Both \nbs-II and \nbs-III exhibit charge density
wave (CDW) transitions\cite{Zettl82v43,Zybtsev17v95}. Recently, Bloodgood
\etal{}\cite{Bloodgood18v6} reported the synthesis of new monoclinic
polymorphs of \nbs, where structural information was obtained via
single crystal x-ray diffraction, electron microscopy, and Raman
spectroscopy studies\cite{Bloodgood18v6}.  The new polymorphs, \nbs-IV
and \nbs-V, were found to crystallize in the space groups $P12_1/c1$
and $P12_1/m1$, respectively.
Presently, first-principles studies of the anharmonic effects that
account for thermal expansion are lacking for these compounds.

The outline of this paper is as follows: \Sec~\ref{subsec:tec}
discusses the concepts of a TEC tensor and \lp{} TECs and the
distinction between them. \Sec~\ref{subsec:formalism} presents the
\Gru{} formalism for the computation of the TEC tensors. For efficient
implementation and a good physical interpretation,
\Sec~\ref{subsec:matrix} shows how the tensorial expression can be
turned into a simple matrix expression. \Sec~\ref{subsec:deform}
discusses the deformations that are to be applied to the crystal for
the calculations of elastic constants and \GP{}s.  The procedure for
elastic constant calculations is outlined in \Sec~\ref{subsec:elastic}
where we observe that all required elastic constants for the TEC
calculations can be obtained from symmetry-preserving deformations.
In \Sec~\ref{subsec:EOM}, we describe how the \lp{} TECs may be
obtained from a TEC tensor by solving a set of coupled differential
equations. The reverse of this process is presented in
Appendix~\ref{sec:exp-tensor}. The results are presented in
\Sec~\ref{sec:results}. Finally \Sec~\ref{sec:summary} contains the
conclusions.

\section{Methodology}

\subsection{TEC tensor versus \lp{} TECs}
\label{subsec:tec}
Thermal expansion of a crystal can generally be described by the
thermal expansion coefficient (TEC) tensor with components
$\alpha_{ij}(T)$ that depend on temperature $T$.  When the temperature
changes, the crystal will change its shape according to a strain
tensor $\epsilon_{ij}$ induced by the TEC tensor where
\begin{equation}
\frac{d\epsilon_{ij}}{dT} = \alpha_{ij} (T)
\end{equation}
In three dimensions, the most general $\alpha_{ij}(T)$ is a $3\times3$
symmetric tensor with six independent parameters.  For a cubic
crystal, it turns out that $ \alpha_{11}(T) = \alpha_{22}(T) =
\alpha_{33}(T) $ while all crossed terms $\alpha_{ij}(T) = 0$, $i\ne
j$, which means that only one independent parameter is needed to
describe the TEC.  Also it does not matter how the cubic crystal is
oriented in space since the expansion is isotropic.  Specifically, the
TEC tensor components $\alpha_{ii}(T)$ are all equal for $i=1,2,3$,
and they coincide with the meaning of the $a$ \lp{} TEC, denoted by
$\alpha_a(T) = \frac{1}{a(T)} \td{a(T)}{T}$ for the cubic crystal 
where $a$ is the lattice constant.  In
another example, the hexagonal crystal is described by two independent
TEC tensor components $\alpha_a(T) = \alpha_{11}(T) = \alpha_{22}(T)
$, and $\alpha_c(T) = \alpha_{33}(T)$ where we must tacitly assume
that the $c$ axis is pointing in the $z$ Cartesian direction.
However, for a low-symmetry monoclinic crystal (with lattice parameters
$a\neq b\neq c$ and $\beta \neq 90^\circ$), the matter becomes
slightly more complicated.  It is usually convenient to choose a
crystal orientation where the $b$ lattice parameter is parallel to the
$y$ Cartesian direction (the so-called `unique axis $b$' in
Ref.~[\onlinecite{ITtable06-book}]).  Symmetry considerations and
Neumann's principle dictates that the TEC tensor of a monoclinic
crystal is of the following form\cite{Schlenker75v60,Nye85-book}
\begin{equation}
\begin{pmatrix} 
\alpha_{11} & 0 & \alpha_{31}\\
0 & \alpha_{22} & 0\\
\alpha_{31} & 0 & \alpha_{33}
\end{pmatrix}
\label{eq:TECtensor}
\end{equation}
Here, there are four independent components of the TEC tensor.
Similar to the basic definition of the \lp{} TEC $\alpha_a(T)$ for a cubic crystal, we define
a \lp{} TEC of a monoclinic crystal according to
\be
\alpha_\ell(T) = \frac{1}{\ell(T)}\frac{d\ell(T)}{dT}
\label{eq:LPTEC}
\ee
where $\ell $ equals to one of the monoclinic lattice parameters $a$,
$b$, $c$, and $\beta$.  Since the $b$ axis is unique by choice,
the $\alpha_{22}(T)$ component is decoupled from the other three
components and coincides with the $b$ \lp{} TEC, where $ \alpha_b(T)
= \alpha_{22}(T) $.  With a `unique axis b' choice, it is still necessary to
clearly state the crystal orientation before reporting the other
three TEC tensor components $\alpha_{11}(T)$, $\alpha_{33}(T)$, and
$\alpha_{31}(T)$.  Unless the $a$ lattice parameter happens to be
aligned along the $x$ Cartesian axis, $\alpha_{11}(T)$ is not same as
the \lp{} TEC for $a$.  It is then apparent that the temperature
dependence of $a$, $c$, and $\beta$ collectively depend on
$\alpha_{11}(T)$, $\alpha_{33}(T)$, and $\alpha_{31}(T)$, and vice
versa.  In \Sec~\ref{subsec:EOM} and Appendix~\ref{sec:exp-tensor} we
derive the relations between these two TEC descriptions.  Throughout
this work, we assume the orientation of a monoclinic crystal in
equilibrium at $T=0$~K is as follows:
\bea
&& \VEC{a} = \azero \VEC{i},   \nonumber
\\ && \VEC{b} = \bzero \VEC{j}, \nonumber
\\ && \VEC{c} = \czero \cos\betazero \VEC{i} + \czero \sin\betazero \VEC{k}.
\label{eq:0Kabc}
\eea
The Cartesian unit vectors are $\VEC{i}$, $\VEC{j}$, and $\VEC{k}$.
It is important to note that for $T > 0$~K, $\VEC{a}$ may have a
nonzero component along $\VEC{k}$ due to presence of the off-diagonal
$\alpha_{31}(T)$ component.

We shall now comment on the nonuniqueness of the choice of $a$, $c$,
and $\beta$ of a monoclinic primitive cell, even within a `unique axis
$b$' choice.  With the specific choice adopted in \Eq~\ref{eq:0Kabc},
we may derive from it a second set of primitive lattice vectors such as
\bea
&& \VEC{a}' = \VEC{a} = \azero \VEC{i},   \nonumber
\\ && \VEC{b}' = \VEC{b} = \bzero \VEC{j}, \nonumber
\\ && \VEC{c}' = \VEC{c} - \VEC{a} = \czero' \cos\betazero' \VEC{i} + \czero' \sin\betazero' \VEC{k},
\label{eq:new0Kabc}
\eea
This will in general have the consequence that $\alpha_{c'} \ne
\alpha_{c}$ and $\alpha_{\beta'} \ne \alpha_{\beta}$.  We shall see
the effect of such a choice on the results in
Section~\ref{sec:results}.

We may calculate the volumetric TEC $\alpha_v$ (with suppressed $T$
dependence for clarity) from
\be
\alpha_v = \left| \begin{array}{ccc} \
1+ \alpha_{11} & 0 & \alpha_{31}  
\\ 0& 1+ \alpha_{22}  & 0 
\\ \alpha_{31} & 0 & 1+ \alpha_{33}  
\end{array} \right| -1 
\ee
The same $\alpha_v$ can also be calculated from the \lp{} TECs where
\be
\alpha_v = \alpha_a + \alpha_b + \alpha_c 
+ \frac{\beta \cos \beta}{\sin \beta} \alpha_{\beta}
\ee
This can be derived from the expression for the volume of 
the primitive cell $\Omega$ where $\Omega = abc \sin\beta$.

\subsection{\Gru{} formalism}
\label{subsec:formalism}
The TEC tensor components $\alpha_{ij}(T)$ for any crystal at temperature 
$T$ may be calculated from first-principles
via a \Gru{} formalism\cite{Gruneisen26v10,Schelling03v68} expressed 
in a tensorial form:
\begin{equation}
\alpha_{ij}(T) = \frac{1}{\Omega} \sum_{k=1}^3 \sum_{l=1}^3 S_{ijkl} I_{kl}(T)
\label{eq:TEC-alphaij}
\end{equation}
The elastic compliance tensor and equilibrium volume of the primitive 
cell are denoted by $S_{ijkl}$
and $\Omega$, respectively.

Central to the \Gru{} formalism are the integrated quantities $I_{ij}(T)$ 
given by
\begin{equation}
I_{ij} (T) = \frac{\Omega}{\left(2\pi\right)^3} \sum_\lambda \int_{\rm BZ} d\VEC{q} \, 
\gamma_{ij,\lambda\VEC{q}} \, c(\nu_{\lambda\VEC{q}},T)
\label{eq:Iij-directsum}
\end{equation}
Here $\nu_{\lambda\VEC{q}}$ denotes the frequency of a phonon mode
with mode index $\lambda$ and wave-vector $\VEC{q}$.  The heat
capacity of a phonon mode of frequency $\nu$ is given by $c(\nu,T) =
\kBol (r/\sinh r)^2$, with $r=h\nu/2\kBol T$, where $h$ and $\kBol$
are the Planck and Boltzmann constants, respectively.  The
bandstructure-like deformation-dependent \GP{} is defined by
\begin{equation}
\gamma_{ij,\lambda\VEC{q}} = - \frac{1}{\nu_{\lambda \VEC{q}}}
\frac{\partial \nu_{\lambda\VEC{q}}}{\partial \epsilon_{ij}}
\label{eq:GP}
\end{equation}
which measures the fractional change in phonon mode frequency with
respect to a small strain $\epsilon_{ij}$ applied to the crystal.
Phonon frequencies may be calculated accurately either from a
density-functional perturbation theory (DFPT)
approach\cite{Baroni01v73} or a supercell force-constant
method\cite{Kresse95v32,Ackland97v9,Alfe09v180}.  The \GP{}s are
calculated based on \Eq~\ref{eq:GP} from phonon frequency shifts using
a central-difference scheme with the help of changes in the dynamical
matrices and a perturbation theory\cite{Gan15v92}.  In this work, we
apply strains of $\epsilon=\pm 0.01$ for all deformations to determine
the \GP{}s.

The integrated quantity $I_{ij}(T)$ in \Eq~\ref{eq:Iij-directsum} is
seen as an integral over the Brillouin zone (BZ) of the product of the
mode-dependent heat capacities and mode-dependent \GP{}s.
It can also be written as an integral over the phonon
frequencies \be I_{ij}(T) = \int_{\nu_{min}}^{\nu_{max}} d\nu
\ \Gamma_{ij}(\nu) c(\nu,T)
\label{eq:Iij-integrate}
\ee where $\nu_{min}$ ($\nu_{max}$) is the minimum (maximum) frequency
in the phonon spectrum.  Here we have introduced a quantity called the
density of the phonon states weighted by the \GP{}s \be
\Gamma_{ij}(\nu) = \frac{\Omega}{\left(2\pi\right)^3} \sum_\lambda
\int_{\rm BZ} d\VEC{q} \ \gamma_{ij,\lambda\VEC{q}} \delta(\nu -
\nu_{\lambda\VEC{q}})
\label{eq:Gammaij}
\ee With the definition in \Eq~\ref{eq:Gammaij}, we turn the
direct summation in \Eq~\ref{eq:Iij-directsum} into an
integral given in \Eq~\ref{eq:Iij-integrate} involving the two
quantities $c(\nu,T)$, which has a form that is easily understood, and
$\Gamma_{ij}(\nu)$, which captures the net effect of anharmonicities of
all phonon modes with the same frequency $\nu$.

\subsection{Compact matrix expression}
\label{subsec:matrix}

We find it convenient to use 
the analogy of the strain-stress relationship\cite{Nye85-book}
$\epsilon_i = S_{ij} \sigma_j$ to cast
\Eq~\ref{eq:TEC-alphaij} into

\begin{equation}
\alpha_i(T) = \frac{1}{\Omega} S_{ij} I_j(T)
\label{eq:TEC-alphai}
\end{equation}
with the prescription
\begin{equation}
  \begin{pmatrix}
       \alpha_{11}(T) &  \alpha_{12}(T) &  \alpha_{31}(T) \\
       \alpha_{12}(T) &  \alpha_{22}(T) &  \alpha_{23}(T) \\
       \alpha_{31}(T) &  \alpha_{23}(T) &  \alpha_{33}(T)
  \end{pmatrix}
\rightarrow
  \begin{pmatrix}
       \alpha_{1}(T) &  \half \alpha_{6}(T) &  \half \alpha_{5}(T) \\
       \half \alpha_{6}(T) &  \alpha_{2}(T) &  \half \alpha_{4}(T) \\
       \half \alpha_{5}(T) &  \half \alpha_{4}(T) &  \alpha_{3}(T)
  \end{pmatrix}
\label{eq:alphaconversion}
\end{equation}
and
\begin{equation}
  \begin{pmatrix}
       I_{11}(T) &  I_{12}(T) &  I_{31}(T) \\
       I_{12}(T) &  I_{22}(T) &  I_{23}(T) \\
       I_{31}(T) &  I_{23}(T) &  I_{33}(T)
  \end{pmatrix}
\rightarrow
  \begin{pmatrix}
       I_{1}(T) &  I_{6}(T) &  I_{5}(T) \\
       I_{6}(T) &  I_{2}(T) &  I_{4}(T) \\
       I_{5}(T) &  I_{4}(T) &  I_{3}(T)
  \end{pmatrix}
\label{eq:Iconversion}
\end{equation}

Now, we need to discuss how to obtain $S_{ij}$ from the elastic constants.
In general, the $6 \times 6$ symmetric elastic constant matrix $C$ with matrix elements $C_{ij}$
consists of $21$ independent matrix elements\cite{Nye85-book}.

For a monoclinic crystal with `a unique axis $b$' choice, 
the matrix $C$ with $13$ independent matrix elements is
\begin{equation}
C = 
  \begin{pmatrix}
       C_{11} &  C_{12}  &  C_{13}  &  0 & C_{15} & 0 \\
       C_{12} &  C_{22}  &  C_{23}  &  0 & C_{25} & 0 \\
       C_{13} &  C_{32}  &  C_{33}  &  0 & C_{35} & 0 \\
       0 &  0 &  0  &  C_{44} & 0 & C_{46} \\
       C_{15} &  C_{25} &  C_{35}  &  0 & C_{56} & 0 \\
       0  &  0&  0  &  C_{46} & 0 &  C_{66} \\
  \end{pmatrix}
\label{eq:monoclinicC}
\end{equation}
We will give more details on the calculation of elastic constants in 
\Sec~\ref{subsec:elastic}.
The elastic compliance matrix $S$ with
matrix elements $S_{ij}$ is simply the inverse of $C$, i.e., $S = C^{-1}$.

\subsection{Crystal deformation}
\label{subsec:deform}
We describe how to perform a deformation to the monoclinic crystal with
lattice parameters $a$, $b$, $c$, and $\beta$. Deformations are required
for the elastic constants and the mode-dependent \GP{}s according to
\Eq~\ref{eq:GP}. 
At $0$~K, we use a matrix $A$ obtained from the primitive lattice vectors (see \Eq~\ref{eq:0Kabc})
\begin{equation}
A = 
  \begin{pmatrix}
       \azero &  0 &  \czero\cos\betazero \\
       0 &  \bzero &  0 \\
       0 &  0  & \czero \sin\betazero
  \end{pmatrix}
\end{equation}
Next we let the deformation matrix\cite{Nye85-book} be
\begin{equation}
E = 
  \begin{pmatrix}
       \epsilon_{11} &  \epsilon_{12} &  \epsilon_{31} \\
       \epsilon_{12} &  \epsilon_{22} &  \epsilon_{23} \\
       \epsilon_{31} &  \epsilon_{23}  & \epsilon_{33}
  \end{pmatrix}
=
  \begin{pmatrix}
       \epsilon_1 &  \half \epsilon_6 &  \half \epsilon_5 \\
       \half \epsilon_6 &  \epsilon_2 &  \half \epsilon_4 \\
       \half \epsilon_5 &  \half \epsilon_4  & \epsilon_3 
  \end{pmatrix}
\label{eq:Ematrix}
\end{equation}
so that the primitive lattice vectors $\VEC{a}$, $\VEC{b}$, and
$\VEC{c}$ after a deformation are described by $A' = (I+E) A $, where
$I$ is the identity matrix.  The full $3\times 3$ matrix $E$ can be
equivalently characterized by a strain vector $\VEC{\epsilon}=
(\epsilon_1, \epsilon_2, \epsilon_3, \epsilon_4, \epsilon_5,
\epsilon_6)$.  In this notation $\VEC{\epsilon} =
(\epsilon_1,0,0,0,0,0)$ describes a uniaxial strain of $\epsilon_1$ in
the $x$ Cartesian direction, while $\VEC{\epsilon}=(0,0,0,0,e_5,0)$
describes a shear strain of $ \epsilon_{31} = \half\epsilon_5$ in the
$x$ and $z$ Cartesian directions.  Since we use the strain vector
$\VEC{\epsilon}$ with single-index entries to describe a deformation,
care should be exercised when dealing with the off-diagonal term
\begin{equation}
\gamma_{31,\lambda \VEC{q}} 
= - \frac{1}{\nu_{\lambda\VEC{q}}} \frac{ \partial \nu_{\lambda\VEC{q}}    }{  \partial \epsilon_{31}} = 
- \frac{2}{\nu_{\lambda\VEC{q}}} 
\frac{ \partial \nu_{\lambda\VEC{q}}}{  \partial \epsilon_{5}} 
\end{equation}
We find it is necessary to define
\be
\gamma_{5,\lambda\VEC{q}} \equiv - \frac{2}{\nu_{\lambda\VEC{q}}} \frac{ \partial \nu_{\lambda\VEC{q}}    }{  \partial \epsilon_{5}}   
\ee
so to make $  \gamma_{5,\lambda\VEC{q}} = \gamma_{31,\lambda\VEC{q}} $.
Note that this is different from the `bare' \GP{}
(which holds
for the diagonal elements $\gamma_1$, $\gamma_2$, and $\gamma_3$)  where
\be
\gamma^b_{5,\lambda\VEC{q}} = - \frac{1}{\nu_{\lambda\VEC{q}}} \frac{ \partial \nu_{\lambda\VEC{q}}    }{  \partial \epsilon_{5}} 
\ee
and hence $\gamma_{5,\lambda\VEC{q}} = 2 \gamma^b_{5,\lambda\VEC{q}}$. 
We then have from \Eq~\ref{eq:Gammaij}
\be
\Gamma_5(\nu) = \Gamma_{31}(\nu) = \frac{\Omega}{\left(2\pi\right)^3} \sum_\lambda \int_{\rm BZ} d\VEC{q}
\     \gamma_{5,\lambda\VEC{q}} \delta(\nu - \nu_{\lambda\VEC{q}})
\ee
and from \Eq~\ref{eq:Iconversion}
\be
I_5(T) = I_{31}(T) = \frac{\Omega}{\left(2\pi\right)^3} \sum_\lambda \int_{\rm BZ} d\VEC{q} \  \gamma_{5,\lambda\VEC{q}} \, c(\nu_{\lambda\VEC{q}},T)
\ee

\subsection{Elastic constant calculation}
\label{subsec:elastic}

In this subsection, we describe how to determine the $13$ elastic 
constants $C_{ij}$ for a monoclinic crystal.
We follow the 
approach in Refs.~\onlinecite{Beckstein01v63,DalCorso16v28}
due to its simplicity.

To further simplify notations we express the elastic constants 
in \Eq~\ref{eq:monoclinicC} as
\begin{equation}
C =   \begin{pmatrix}
       c_{1} &  c_{2}  &  c_{3}  &  0 & c_{4} & 0 \\
       c_{2} &  c_{5}  &  c_{6}  &  0 & c_{7} & 0 \\
       c_{3} &  c_{6}  &  c_{8}  &  0 & c_{9} & 0 \\
       0 &  0 &  0  &  c_{10} & 0 & c_{11} \\
       c_{4} &  c_{7} &  c_{9}  &  0 & c_{12} & 0 \\
       0  &  0&  0  &  c_{11} & 0 &  c_{13} \\
  \end{pmatrix}
\end{equation}

These constants can be obtained by performing thirteen independent
deformation types according to \Table~\ref{tab:deformation}. For each
deformation type, a set of strains $\epsilon$ ranging from $-0.01$ to
$0.01$ is used to obtain an energy $\Delta E/\Omega$ versus strain
$\epsilon$ curve. A quadratic least-square fit is used to obtain the
curvature $k$ for the parabola. Each curvature is in general a linear
combination of a few elastic constants, e.g., for the deformation type
$7$, since $\frac{\Delta E}{\Omega} = \half (c_1 +2c_2 + c_5)
\epsilon^2$ (see \Table~\ref{tab:deformation}), $k_7=\half (c_1 +2c_2
+ c_5)$.  The values of $c_i$, $i=1, 2, \ldots, 13$ can be obtained
from $k_i$ by a matrix inversion.

We note that of all the thirteen deformations, ten of them preserve
the space group of the crystal after a deformation is applied.  These
are the deformation types $1$, $2$, $3$, $5$, $7$, $8$, $9$, $10$,
$11$, and $12$.  Since the effect of temperature does not change the
crystal type (i.e., the crystal remains monoclinic upon heating or
cooling), we expect that $I_4(T)= I_6(T) = 0$, and
\Eq~\ref{eq:TEC-alphai} is thus reduced to
\be
  \begin{pmatrix}
       \alpha_1 \\
       \alpha_2 \\
       \alpha_3 \\
       \alpha_5   
  \end{pmatrix}
=
\frac{1}{\Omega}
  \begin{pmatrix}
       c_1 &  c_2 &  c_3 &  c_4 \\
       c_2 &  c_5 &  c_6 &  c_7 \\
       c_3 &  c_6 &  c_8 &  c_9 \\
       c_4 &  c_7 &  c_9 &  c_{12}
  \end{pmatrix}^{-1}
  \begin{pmatrix}
       I_1 \\
       I_2 \\
       I_3 \\
       I_5 
  \end{pmatrix}
\ee
For a monoclinic crystal, we observe that all ten
deformations needed for the TEC calculation can be chosen to be
strictly space-group preserving. This is important for an accurate
determination of elastic constants because we can maintain the same
number of $\VEC{k}$ points throughout all self-consistent
density-functional theory calculations, which minimizes possible
total-energy fluctuation.  The existence of a minimum yet sufficient
number of symmetry-preserving deformation types to calculate just all
relevant elastic constants for the sole purpose of TEC determination
is also observed in other crystal types: cubic with one uniform
deformation, orthorhombic\cite{Gan15v92} with six deformations,
hexagonal and trigonal with three
deformations\cite{Gan16v94,Gan18v151}, and of course, the triclinic
system with the full 21 deformations for all 21 elastic constants (the
space group of a triclinic crystal is either $P1$ or $P{\overline 1}$).

\begin{table}
\begin{center}
\begin{tabular}{l|l|l}
\hline
Deformation& Strain & $\Delta E/\Omega$ to $O(\epsilon^2)$ \\
type & vector $\VEC{\epsilon}$  & \\
\hline
\hline
1 & $(\epsilon,0,0,0,0,0)$ & $ \frac{1}{2} c_1 \epsilon^2$ \\
2 & $(0,\epsilon,0,0,0,0)$ & $ \frac{1}{2} c_5 \epsilon^2$ \\
3 & $(0,0,\epsilon,0,0,0)$ & $ \frac{1}{2} c_8 \epsilon^2$ \\
4 & $(0,0,0,\epsilon,0,0)$ & $ \frac{1}{2} c_{10} \epsilon^2$ \\
5 & $(0,0,0,0,\epsilon,0)$ & $ \frac{1}{2} c_{12} \epsilon^2$ \\
6 & $(0,0,0,0,0,\epsilon)$ & $ \frac{1}{2} c_{13} \epsilon^2$ \\
7 & $(\epsilon,\epsilon,0,0,0,0)$ & $ \frac{1}{2} (c_1 + 2 c_2 + c_{5} ) \epsilon^2$ \\
8 & $(\epsilon,0, \epsilon,0,0,0)$ & $ \frac{1}{2} (c_1 + 2 c_3 + c_{8} ) \epsilon^2$ \\
9 & $(0,\epsilon,\epsilon,0,0,0)$ & $ \frac{1}{2} (c_5 + 2 c_6 + c_{8} ) \epsilon^2$ \\
10 & $(\epsilon,0,0,0,\epsilon,0)$ & $ \frac{1}{2} (c_1 + 2 c_4 + c_{12} ) \epsilon^2$ \\
11 & $(0,\epsilon,0,0,\epsilon,0)$ & $ \frac{1}{2} (c_5 + 2 c_7 + c_{12} ) \epsilon^2$ \\
12 & $(0,0,\epsilon,0,\epsilon,0)$ & $ \frac{1}{2} (c_8 + 2 c_9 + c_{12} ) \epsilon^2$ \\
13 & $(0,0,0,\epsilon,0,\epsilon)$ & $ \frac{1}{2} (c_{10} + 2 c_{11} + c_{13} ) \epsilon^2$\\
\hline
\hline
\end{tabular}
\end{center}
\caption{
Thirteen deformation types for the calculation of all thirteen elastic
constant for a monoclinic crystal.  The strain vector $\VEC{\epsilon}$
defines a deformation applied to the crystal in equilibrium according
to \Eq~\ref{eq:Ematrix}.  The third column describes the increase of
the total energy $\Delta E$ as a quadratic dependence on $\epsilon$.
}
\label{tab:deformation}
\end{table}

\subsection{Calculation of the \lp{} TECs}
\label{subsec:EOM}

Having obtained the TEC tensor $\alpha_i(T)$, we still need a recipe
to find the \lp{} TECs as defined in \Eq~\ref{eq:LPTEC}.  Since $b$ is
not coupled to $a$, $c$, and $\beta$, we have immediately $\alpha_b(T)
= \alpha_2(T)$.  To find the other three \lp{} TECs $\alpha_a$,
$\alpha_c$, and $\alpha_{\beta}$, we need to know how the lattice
parameters $a, c, \beta $ vary with $T$ and then use a simple
numerical difference scheme (such as forward difference) to extract
$\alpha_{\ell}$ according to \Eq~\ref{eq:LPTEC}.  
Here we need
to derive a set of differential equations for numerical integration.  We
assume the primitive lattice vectors at temperature $T$ are
$\VEC{a}(T)= a_1(T) \VEC{i} + a_3(T) \VEC{k}$, $\VEC{b}(T)=b_2(T)
\VEC{j}$, and $\VEC{c}(T)=c_1(T) \VEC{i} + c_3(T)\VEC{k}$.  At
$T=0$~K, we have the initial conditions $\VEC{a}(0) = (\azero, 0,0)$,
$\VEC{b}(0) = (0, \bzero,0)$, $\VEC{c}(0) = (\czero \cos\betazero, 0,
\czero \sin\betazero)$.
From \Eq~\ref{eq:alphaconversion} we let
\be
{\widetilde \alpha}(T) = 
  \begin{pmatrix}
       \alpha_{1}(T) &  \half \alpha_{6}(T) &  \half \alpha_{5}(T) \\
       \half \alpha_{6}(T) &  \alpha_{2}(T) &  \half \alpha_{4}(T) \\
       \half \alpha_{5}(T) &  \half \alpha_{4}(T) &  \alpha_{3}(T)
  \end{pmatrix}
\ee
When $T$ changes to $T'=T + \Delta T$, we have
\begin{equation}
  \begin{pmatrix}
       a_1(T') &  0 &  c_1(T') \\
       0 & b_2(T') & 0\\
      a_3(T') & 0& c_3(T')
  \end{pmatrix}
=
D
  \begin{pmatrix}
       a_1(T) &  0 &  c_1(T) \\
       0 & b_2(T) & 0\\
      a_3(T) & 0& c_3(T)
  \end{pmatrix}
\end{equation}
where the deformation matrix $D$ is given by
\be
D = I + {\widetilde \alpha}(T) \Delta T
\ee
As $\Delta T$ goes to zero, we obtain five differential equations:
\begin{eqnarray}
\frac{d}{dT} a_1(T) &=& \alpha_1(T) a_1(T) + \half \alpha_5(T) a_3(T) 
\label{eq:a1prime}
\\ \frac{d}{dT} a_3(T) & = & \half \alpha_5(T) a_1(T)+ \alpha_3(T) a_3(T)
\label{eq:a3prime}
\\ \frac{d}{dT} b_2(T) & = & \alpha_2(T) b_2(T)
\label{eq:b2prime}
\\ \frac{d}{dT} c_1(T)& = & \alpha_1(T) c_1(T) + \half \alpha_5(T) c_3(T)
\\ \frac{d}{dT} c_3(T)& = & \half \alpha_5(T) c_1(T) + \alpha_3(T) c_3(T)
\end{eqnarray}

Since $b(T) = | \VEC{b}(T)| = b_2(T)$, we have immediately $\alpha_b(T) = \alpha_2(T)$ from \Eq~\ref{eq:b2prime}, which is
expected since the primitive lattice vector $\VEC{b}$ is not 
coupled to $\VEC{a} $ and $\VEC{c}$.
With the initial conditions $a_1(0) = \azero$, $a_3(0) = 0$, 
$c_1(0) = \czero \cos\betazero$, and $c_3 (0) = \czero \sin\betazero$,
we use a robust fourth-order Runge-Kutta integration scheme to numerically integrate the
differential equations. From $a(T) = \sqrt{a_1(T)^2 + a_3(T)^2}$, and
$c(T) = \sqrt{ c_1(T)^2 + c_3(T)^2}$, we obtain the 
\lp{} TECs $\alpha_a$ and $\beta_c$, and $\alpha_{\beta}$.
Now we have outlined a procedure to obtain the \lp{} TECs from 
the TEC tensor. In Appendix~\ref{sec:exp-tensor}, we outline
a procedure to reverse the process, which may be more applicable in 
experimental settings since \lp{} TECs are usually measured first.

\section{Results and discussions}
\label{sec:results}

We carry out density-functional theory (DFT) calculations within the
local-density approximation, with projector
augmented-wave (PAW) pseudopotentials as
implemented in the Vienna Ab initio Simulation Package
(VASP)\cite{Kresse96v6} on \nbs{}-IV.  The initial
structure of \nbs{}-IV is taken from Ref.~[\onlinecite{Bloodgood18v6}]
where $\beta=90.12^\circ$, $a=6.752$, $b=4.974$, and
$c=18.132$~\AA. This corresponds to the $P12_1/c1$ setting of the
space group~\#14.  The primitive cell consists of $32$ atoms, $8$
of which are inequivalent.  A large cutoff energy of $400$~eV is used
throughout this work. A $12\times12\times4$ Monkhorst-Pack mesh is
used for the $\VEC{k}$-point sampling for the relaxation of the
lattice parameters of the primitive cell and atomic positions as well
as for the calculations of elastic constants.  Atomic relaxation is
stopped when the maximum force on all atoms is less than
$10^{-3}$~eV/\AA.  To fully optimize the monoclinic lattice
parameters, we successively carry out deformations of types $1$, $2$,
$3$, and $5$ in \Table~\ref{tab:deformation}. For each type of
deformation, we vary the strain $\epsilon$ from $-0.01$ to $0.01$ to
locate the minimum in the energy $E$ versus $\epsilon$ curve.  The
predicted minimum in the $E$-$\epsilon$ curve allows us to adjust the
lattice parameters before a deformation of another type is carried
out. At the end of the procedure, we obtain $\beta=89.98^\circ$,
$a=6.673$, $b=4.870$, and $c=17.837$~\AA, which compare favorably with
the experimental results\cite{Bloodgood18v6}.

We then perform the calculations for the elastic constants as outlined
in \Sec~\ref{subsec:elastic}. For each type of deformation, we also
vary $\epsilon$ from $-0.01$ to $0.01$.  The elastic constants
$C_{ij}$ of \nbsfour{} are $183.64$, $20.49$, $26.32$, $-0.58$,
$158.08$, $7.99$, $0.81$, $47.79$, $1.21$, $7.20$, $0.13$, $23.47$,
and $33.38$~GPa, for $ij= 11$, $12$, $13$, $15$, $22$, $23$, $25$,
$33$, $35$, $44$, $46$, $55$, and $66$, respectively.  Subsequently,
we obtain the elastic compliance matrix $S$ (in $10^{-3}$/GPa) where
\be
S =
  \begin{pmatrix}
    5.97 & -0.61 & -3.20 &  0.00 &  0.33 &  0.00 \\
   -0.61 &  6.44 & -0.73 &  0.00 & -0.20 &  0.00 \\
   -3.20 & -0.73 & 22.84 &  0.00 & -1.23 &  0.00 \\
    0.00 &  0.00 &  0.00 &138.90 &  0.00 & -0.54 \\
    0.33 & -0.20 & -1.23 &  0.00 & 42.69 &  0.00 \\
    0.00 &  0.00 &  0.00 & -0.54 &  0.00 & 29.96
  \end{pmatrix} 
\label{eq:Sij}
\ee

Next we perform the phonon calculations using the supercell
force-constant
method\cite{Gan10v49,Liu14v16}.
We use a $3\times 3 \times 1$ supercell (with $288$ atoms) with a
$4\times4\times4$ Monkhorst-Pack mesh for the $\VEC{k}$-point
sampling. A small $0.015$~\AA{} atomic displacement is used to induce
forces on all atoms in the supercell.  Due to a relatively low
symmetry of the space group \#14, a total of $48$ independent
self-consistent DFT calculations are to be performed for a complete
set of phonon calculations.

\begin{figure}[htbp]
\centering\includegraphics[width=8.2cm,clip]{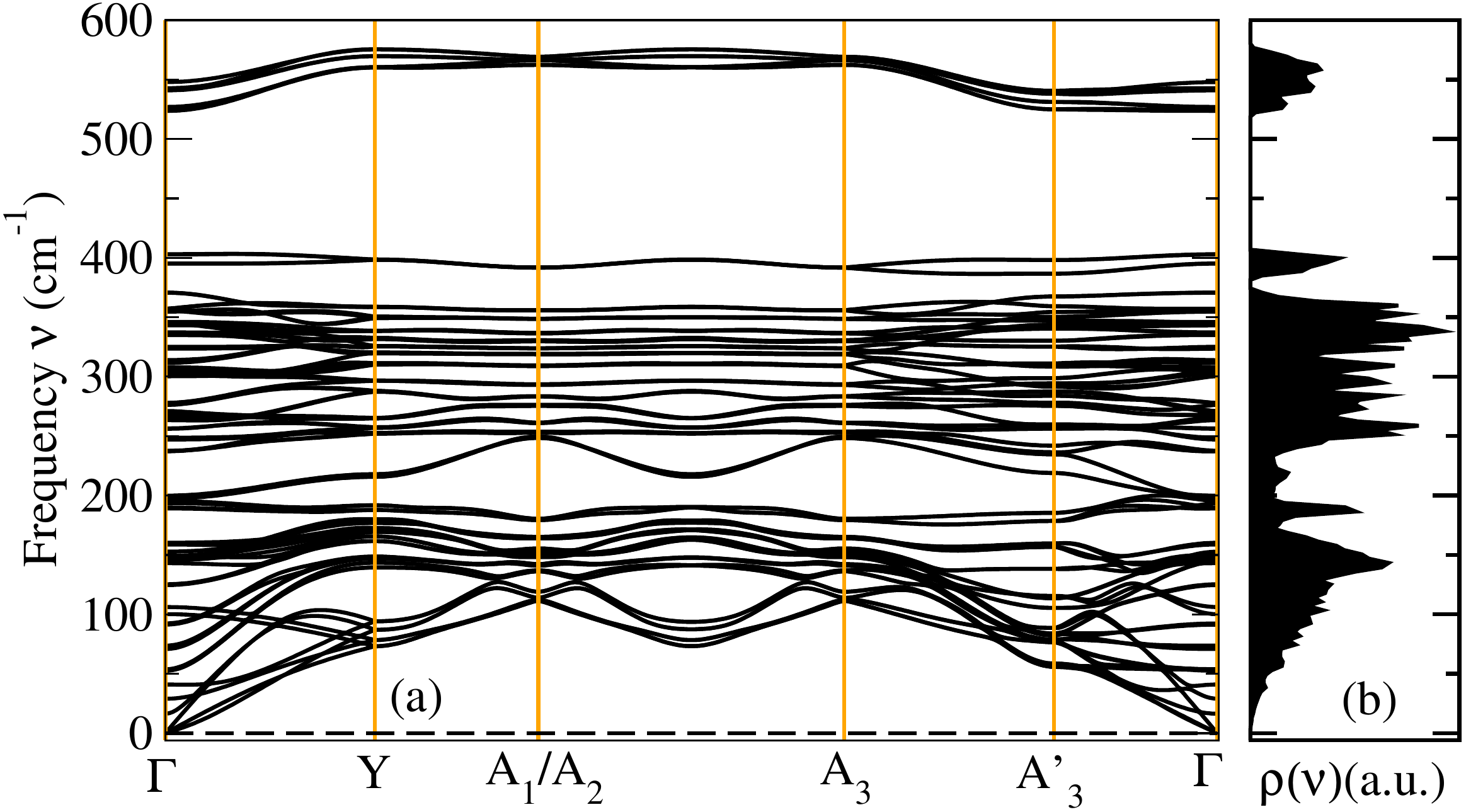}
\caption{(a) The phonon dispersions of the \nbsfour{} along the high-symmetry directions.
The selected $\VEC{q}$ points are (in $\VEC{b}_1$, $\VEC{b}_2$, and
$\VEC{b}_3$): $\Gamma = [0,0,0]$, $Y=[0,\half,0]$, $A_1 =
[r',\half,r]$, $A_2=[s,\half,s']$, $A_3 = [-s, \half,r]$, $A'_3 =
[-s,0,r]$.  Here $r = (a-c\cos\beta)/(2a \sin^2 \beta)$, $r' = \half +
(ar \cos\beta)/c$, $s = (c-a\cos\beta)/(2c \sin^2 \beta)$, and $s' =
\half + (cs \cos\beta)/a$.  (b) The phonon density of states
$\rho(\nu)$, obtained with a $\VEC{q}$-point sampling by a $30 \times 30
\times 10$ mesh.
}
\label{fig:PD}
\end{figure}

For the equilibrium structure, we plot the phonon dispersions along a
few high-symmetry directions as shown in \Fig~\ref{fig:PD}(a).  The
phonon spectrum has a highest frequency of $577$~\invcm{} and a phonon
gap between $407$ and $519$~\invcm{}.  The phonon density of states
$\rho(\nu) = \frac{\Omega}{(2\pi)^3}\sum_{\lambda} \int_{\rm BZ}
d\VEC{q} \ \delta(\nu - \nu_{\lambda\VEC{q}}) $ shown in
\Fig~\ref{fig:PD}(b) does not exhibit imaginary modes, implying that 
\nbsfour{} is stable. 

\begin{figure}[htbp]
\centering\includegraphics[width=8.2cm,clip]{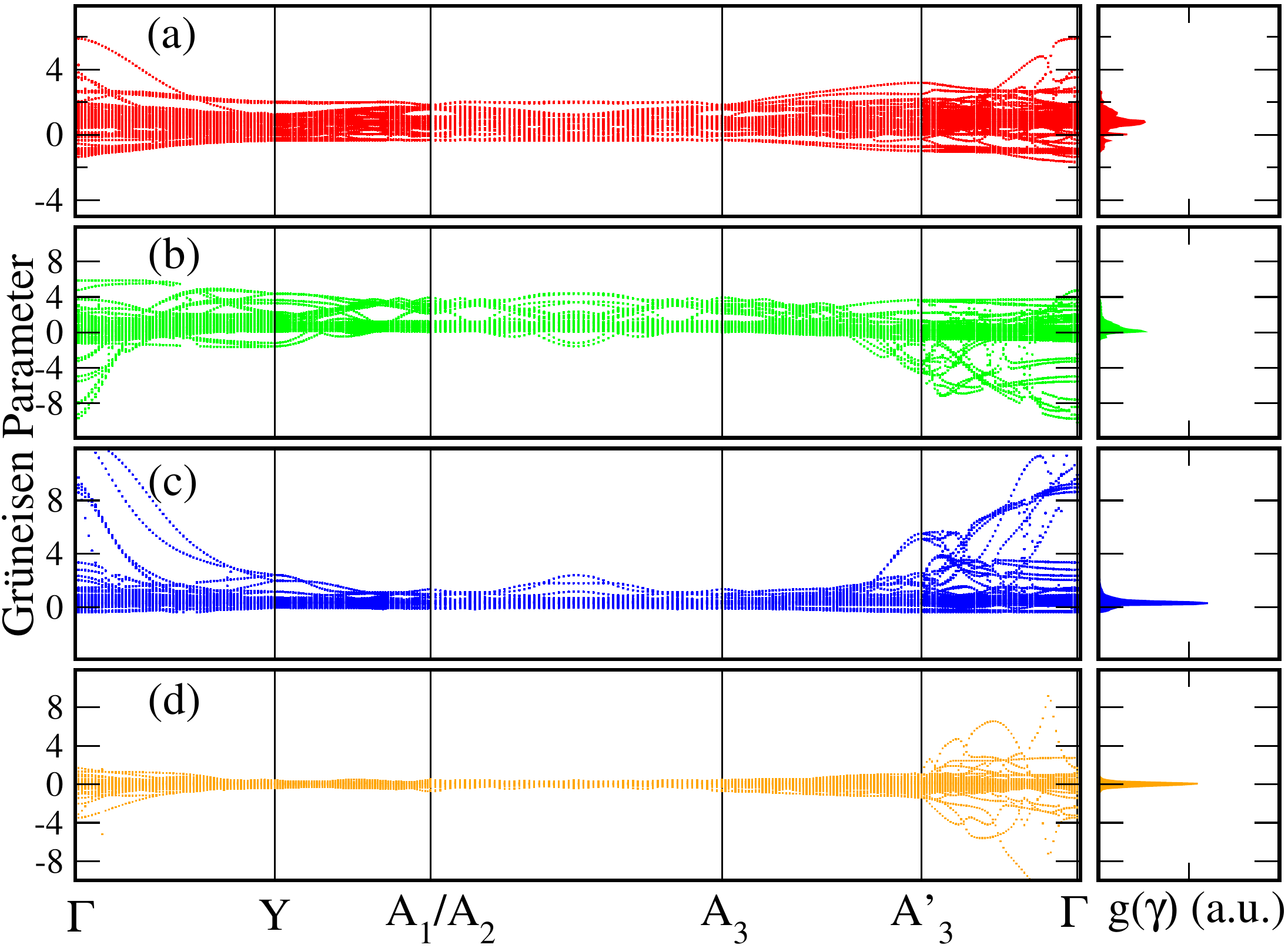}
\caption{The \GP{}s $\gamma_{i,\lambda\VEC{q}}$ of \nbsfour{} along the high-symmetry directions 
for (a) type $1$ ($x$ uniaxial),
(b) type $2$ ($y$ uniaxial), (c) type $3$ ($z$ uniaxial), 
and (d) type $5$ ($xz$ strain).
The selected $\VEC{q}$ points are the same as in \Fig~\ref{fig:PD}.
The right panels are the corresponding densities of \GP{}s $g_i(\gamma)$.
}
\label{fig:GP}
\end{figure}

To obtain all the necessary \GP{}s for calculation of the TEC
tensor, we apply deformations of types $1$, $2$, $3$, and $5$ in
\Table~\ref{tab:deformation} with $\epsilon=\pm 0.01$.  This entails a
total of eight sets of phonon calculations, with each set involves
$48$ independent self-consistent DFT calculations on a large $3 \times
3 \times 1$ supercell. From the frequency changes, we obtain the \GP{}s
as shown in \Fig~\ref{fig:GP}(a)-(d). We see that \GP{}s with large
magnitudes tend to be found near $\Gamma$, where low-frequency phonons
are found.  We calculate the density of \GP{}s according to
$g_i(\gamma) = \frac{\Omega}{(2\pi)^3} \sum_{\lambda} \int_{\rm BZ}
d\VEC{q} \ \delta(\gamma - \gamma_{i,\lambda\VEC{q}}) $ for each $i$,
and the right panels of \Fig~\ref{fig:GP} indicate that most \GP{}s
are concentrated in the $-2$ to $2$ range.

\begin{figure}[htbp]
\centering\includegraphics[width=8.2cm,clip]{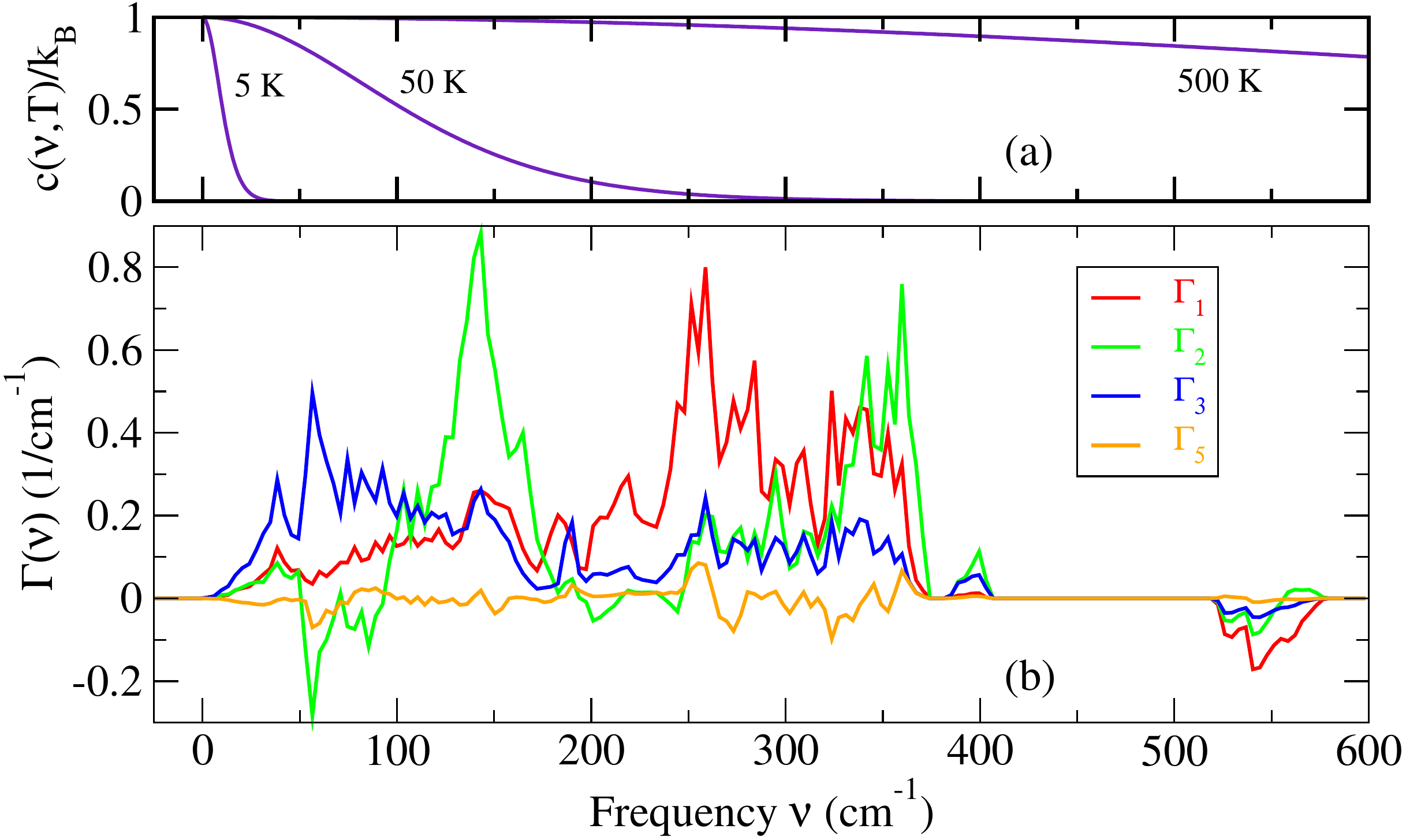}
\caption{
(a) The heat capacity
$c(\nu,T)$ as a function of phonon frequency $\nu$ at $5$, $50$, $500$~K.
(b) $\Gamma_i(\nu)$, the phonon density of states weighted by the \GP{}s
as a function of frequency $\nu$ 
corresponding to type $i$ deformation, for $i=1,2,3,5$.
}
\label{fig:Gamma}
\end{figure}

We then evaluate $\Gamma_i(\nu)$, the phonon density of states
weighted by the \GP{}s, for $i=1,2,3,5$.  A $30\times30\times10$ mesh
is used for the $\VEC{q}$-point sampling.  \Fig~\ref{fig:Gamma}(b)
shows very high anisotropies of the \GP{}s associated with different
deformations.  There is a gap between $407$ to $519$~\invcm{} in all
$\Gamma_i(\nu)$, which is similar to that of the phonon density of
states shown in \Fig~\ref{fig:PD}(b).  The results show that high-frequency
modes (around $550$~\invcm{}) are dominated by negative \GP{}s.  The
dependency of the heat capacity on temperature shown in
\Fig~\ref{fig:Gamma}(a) tells us that these modes contribute to
$I_i(T)$ (via \Eq~\ref{eq:Iij-integrate}) only at high temperatures.
The deformation due to an $xz$ shear strain gives $\Gamma_5(\nu)$
which is only slightly anharmonic relative to that of the diagonal
deformations (i.e., $\Gamma_1(\nu)$, $\Gamma_2(\nu)$, and
$\Gamma_3(\nu)$). These diagonal deformations are mostly positive,
except for $\Gamma_2(\nu)$ in the range of $50$ to $75$~\invcm{}.

\begin{figure}[htbp]
\centering\includegraphics[width=8.2cm,clip]{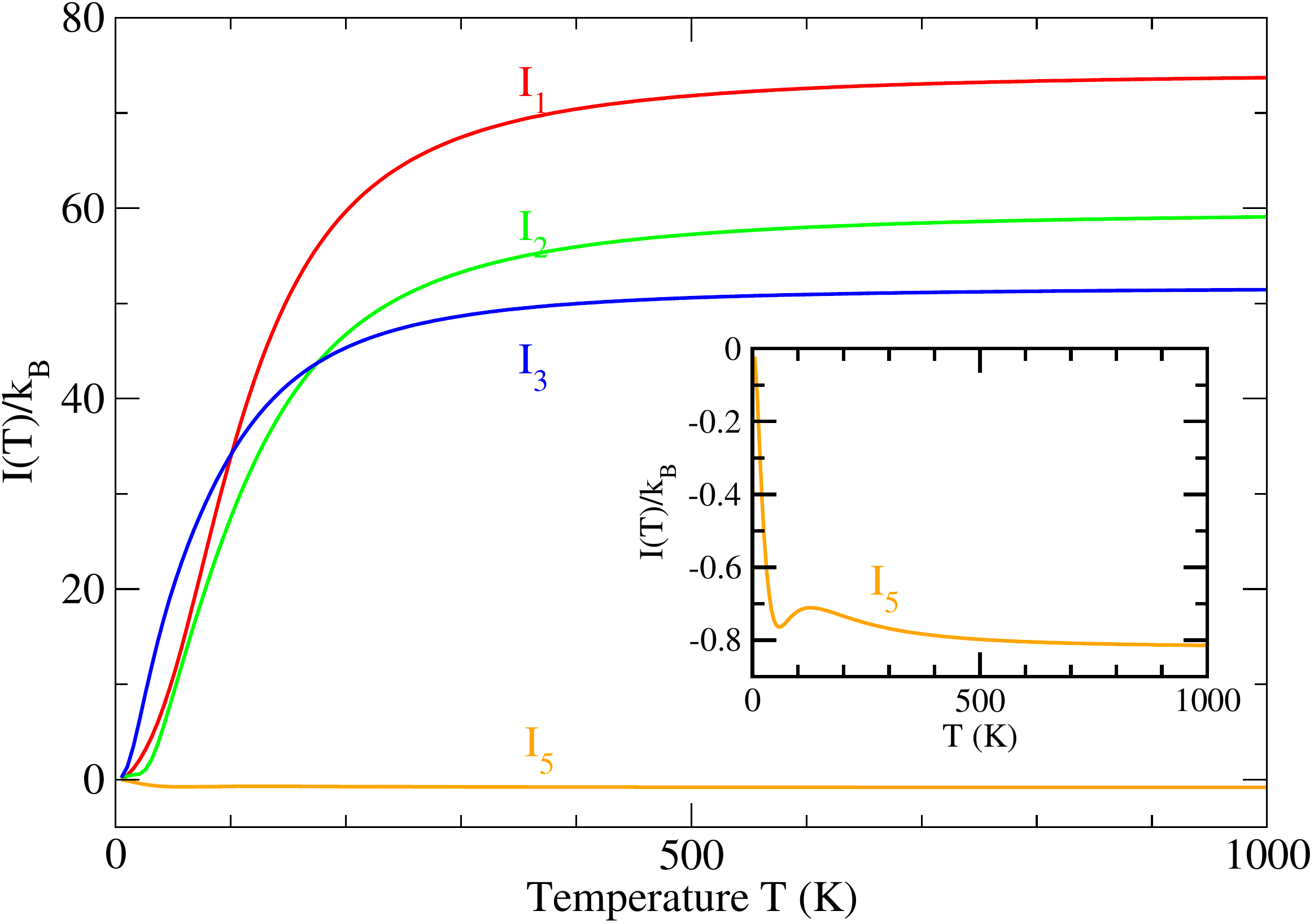}
\caption{
Integrated quantities $I_i(T)$ for different deformations.
The insert shows the detailed variations of $I_5(T)$.
}
\label{fig:Ii}
\end{figure}

The integrated quantities $I_{i}(T)$ for $i=1$, $2$, $3$, and $5$ are
shown in \Fig~\ref{fig:Ii} may be obtained through either a direct
summation in the $\VEC{q}$-space according to
\Eq~\ref{eq:Iij-directsum} or through an integral over the frequencies
according to \Eq~\ref{eq:Iij-integrate}.  It is seen that $I_1(T)$ is
larger than $I_2(T)$ and $I_3(T)$ at high temperatures above $100$~K,
which is consistent with the data in \Fig~\ref{fig:Gamma}(b). It is
seen that $I_5(T)$ is indeed small compared to the diagonal quantities
$I_1(T)$, $I_2(T)$, and $I_3(T)$ over the entire temperature range,
which is again consistent with the small values of $\Gamma_5(\nu)$
shown in \Fig~\ref{fig:Gamma}(b).

\begin{figure}[htbp]
\centering\includegraphics[width=8.2cm,clip]{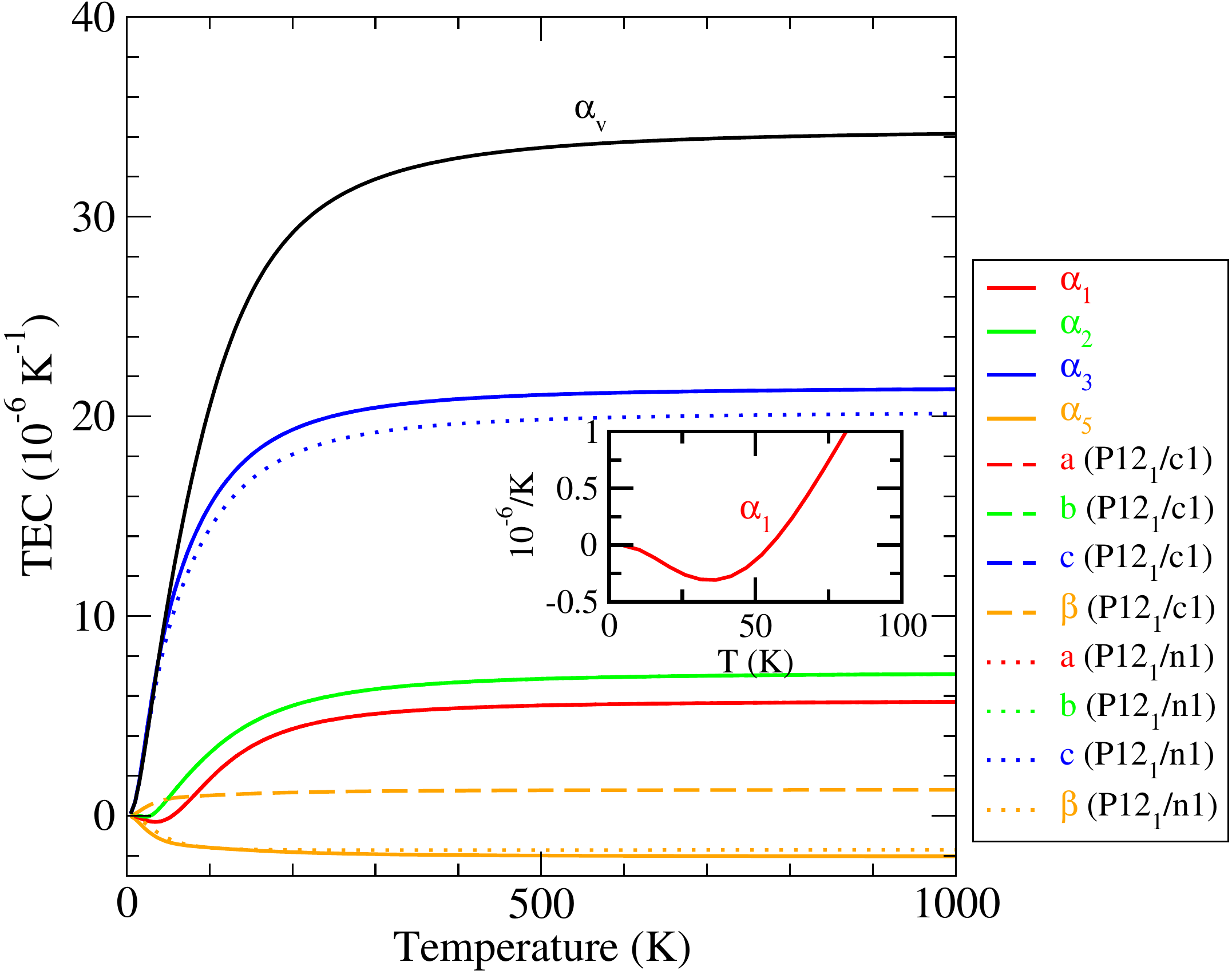}
\caption{The TEC tensor components $\alpha_i$ for $i=1,2,3,5$, the lattice
parameter TECs for $P12_1/c1$ and $P12_1/n1$ settings are shown 
as functions of temperature.
Note that $\alpha_2(T) = \alpha_{\beta}(T) = \alpha_{\beta'}(T)$.
$\alpha_{a}(T) = \alpha_{a'}(T) \approx \alpha_1(T)$. 
$\alpha_{c}(T) \approx \alpha_3(T)$.
}
\label{fig:TEC}
\end{figure}

The TEC tensor components $\alpha_{i}$, $i=1,2,3,5$ are displayed as
solid lines in \Fig~\ref{fig:TEC}.  It is somewhat surprising that
$\alpha_1(T)$ is the smallest among all diagonal TEC tensor
components, despite the fact that $I_1(T)$ is the dominant diagonal
component, at least for $T > 100$~K as shown in \Fig~\ref{fig:Ii}.  We
can rationalize this result as follows. According to
\Eq~\ref{eq:TEC-alphai}, $\alpha_1(T)$ is a linear combination of
$I_i(T)$, weighted by coefficients taken from the first row of the
elastic compliance matrix $S$ in \Eq~\ref{eq:Sij}.  Since $S_{13} =
-3.20\times 10^{-3}$/GPa is negative and large in magnitude compared
to $S_{11} = 5.97\times 10^{-3}$/GPa, and $I_1(T)$ and $I_3(T)$
are comparable in magnitudes, this makes $\alpha_1(T)$ the smallest
among all diagonal $\alpha_i(T)$. It should be noted that
$\alpha_1(T)$ is negative between $0$ and $50$~K (shown in the inset
of \Fig~\ref{fig:TEC}) partly because $I_3(T)$ is dominant for
temperatures below $100$~K.  On the other hand, the dominant value of
$S_{33} = 22.84\times 10^{-3}$/GPa over $S_{31}$ and $S_{32}$ makes
$\alpha_3(T)$ the dominant diagonal TEC tensor component.  
Thus, we conclude that the value of a TEC depends on
not only on the anharmonic effects measured by the
\GP{}s, as commonly believed, but also on the mechanical properties 
such as the elastic constants (or
equivalently, the elastic compliance matrix).  The volumetric TEC of
\nbsfour{} is $ \alpha_v = 34 \times 10^{-6}$~K$^{-1}$ at high
temperatures.

Next, we turn to the discussion of the \lp{} TECs for the crystal in a
$P12_1/c1$ setting with $\beta=89.98^\circ$.  Firstly, $\alpha_b(T) =
\alpha_2(T)$ by the choice of the unique axis.  From
\Fig~\ref{fig:TEC}, we find that $\alpha_a(T) $ is indistinguishable
from $\alpha_1(T)$ since the primitive lattice vector $\VEC{a} $ is
parallel to the $x$ axis at $0$~K and it is always almost parallel to
the $x$ axis even at higher $T$ due to small values of $\alpha_5(T)$.
For this setting in which $\beta$ is very close to $90^\circ$, the primitive
lattice vector $\VEC{c}$ remains almost parallel to the $z$-axis at all
temperatures, causing $\alpha_c(T)$ to be indistinguishable from
$\alpha_3(T)$.  We find that $\alpha_{\beta}(T) $ can be approximated
very well (not shown) by $-\frac{2}{\pi} \alpha_5(T)$, since $\beta(T)$
is always very close to $\half \pi$ and that $ \beta = \half \pi -
\epsilon_5 $ (the physical meaning of $\epsilon_5$ is the decrease in
angle between two elements in the $x$ and $z$ directions, see, e.g.,
Ref.~\onlinecite{Nye85-book}), and hence $ \alpha_{\beta} =
\frac{1}{\beta} \td{\beta}{T} = -\frac{1}{\beta} \alpha_5(T) \approx
-\frac{2}{\pi} \alpha_5(T)$.

We discuss next the \lp{} TECs of a new primitive cell derived from
the $P12_1/c1$ setting with $\beta=89.98^\circ$, using the
transformation rule as shown in \Eq~\ref{eq:new0Kabc}.  It turns out
that this corresponds to a $P12_1/n1$ setting\cite{Nespolo16v72} with $\beta' =
110.50^\circ$, $a' = a = 6.673$, $b' = b = 4.870$, $c' =
19.042$~\AA{}.  Since this setting also describes the same {\em
 unrotated} crystal as the $P12_1/c1$ setting, the TEC tensor components
remain the same. For this new $P12_1/n1$ setting, we evolve the monoclinic lattice parameters as previously performed for the $P12_1/c1$ setting.  It is not surprising to find
that $\alpha_{b'}(T) = \alpha_b(T) = \alpha_2(T) $. In addition, $\alpha_{a'}(T) =
\alpha_a(T)$, both of which are essentially the same as $\alpha_1(T)$.
However, we do find that $\alpha_{c'}(T)$ is somewhat different from $\alpha_{3}(T)$
as a result of the primitive lattice vector $\VEC{c}'$ being oriented away from
$z$ axis. Finally, $\alpha_{\beta'}(T) $ has an opposite sign compared to
$\alpha_{\beta}(T)$ and agrees fortuitously with $\alpha_5(T)$.

\section{Summary}
\label{sec:summary}

With an extension of the \Gru{} formalism,
we have, for the first time, studied the thermal expansion properties
of a monoclinic phase of niobium trisulfide (\nbs{}) using density-functional theory calculations.
Large
anisotropies in the thermal expansion coefficient (TEC) tensor components in the $x$, $y$, $z$
directions have been found. A negative TEC tensor component $\alpha_{11}$  is found at low
temperatures, the occurence of which is mainly due to the mechanical property 
characterized by the elastic compliance. This is somewhat surprising
since a negative TEC is usually associated with the occurence of many phonon modes with
negative \GP{}s.
We expect that the \GP{}s generated in this work could be used to provide
a good estimate of the thermal conductivity based on a Slack's
expression\cite{Toher14v90,Slack79v34,Blanco04v57}. The obtained
\GP{}s can be further used to extract third-order interatomic force constants
for the determination of thermal conductivity by solving the Boltzmann
transport equation\cite{Li14v185,Lindsay16v20}.
Our developed method can be applied to any crystal classes, which opens the door
to studying many important materials at the level of accurate density-functional theory.

\section{Acknowledgments}
We thank the National Supercomputing Center, Singapore (NSCC) and
A*STAR Computational Resource Center, Singapore (ACRC) for computing
resources.  This work is supported in part by RIE2020 Advanced
Manufacturing and Engineering (AME) Programmatic Grant No A1898b0043.

\bibliographystyle{plain}
%
\bibliographystyle{apsrev4-1}

\appendix

\section{Obtaining the TEC tensor from the \lp{} TECs}
\label{sec:exp-tensor}

Given the \lp{} TECs $\alpha_a(T)$ $\alpha_b(T)$, $\alpha_c(T)$, and
$\alpha_{\beta}(T)$, and temperature dependence of the primitive
lattice vectors $\VEC{a}$, $\VEC{b}$, and $\VEC{c}$ where $ \VEC{a}(T)
= a_{1}(T) \VEC{i} + a_{3}(T) \VEC{k}$, $ \VEC{b}(T) = b_2(T)
\VEC{j}$, and $\VEC{c} = c_1(T) \VEC{i} + c_3(T) \VEC{k}$, we seek to
obtain the TEC tensor components $\alpha_i(T)$, for $i = 1,2,3,5$.  As
mentioned before, we have immediately $\alpha_2(T) = \alpha_b(T)$. In
the following we present a method to calculate $\alpha_i(T)$ for
$i=1,3,5$.

Consider $a(T) = \sqrt{a_1(T)^2 + a_3(T)^2}$, we have
\be
a' = \frac{a_1 a_1' + a_3 a_3' }{a}
\label{eq:aprime}
\ee
where $y'$ means the derivative of $y$ with respect to $T$.  We have
suppressed the explicit dependence of all quantities on temperature in
\Eq~\ref{eq:aprime} for clarity.  Replacing $a_1'$ and $a_3'$ using
\Eqs~\ref{eq:a1prime} and \ref{eq:a3prime} we have
\be
a_1^2 \alpha_1 + a_3^2 \alpha_3 +  a_1 a_3 \alpha_5 = a^2 \alpha_a
\ee
The three unknowns $\alpha_1$, $\alpha_3$, and $ \alpha_5$ are seen to
appear in a linear equation.  Therefore we need two other linear
equations to enable a matrix inversion.  A second relation is obtained
by considering $c(T) = \sqrt{c_1(T)^2 + c_3(T)^2} $ that gives
\be
c_1^2 \alpha_1 + c_3^2 \alpha_3  + c_1 c_3 \alpha_5 = c^2 \alpha_c
\ee
The third relation is obtained by differentiating 
$\VEC{a}\cdot \VEC{c} = a_1 c_1 + a_3 c_3 =   a c \cos\beta $
with respect to $T$.
Collecting all three relations, we have
\be
  \begin{pmatrix}
       a_1^2     &  a_3^2     &  a_1 a_3              \\
       c_1^2     &  c_3^2     &  c_1 c_3              \\
       2a_1 c_1  &  2a_3 c_3  &  a_1 c_3 + a_3 c_1  
  \end{pmatrix}
 \begin{pmatrix}
       \alpha_1 \\
       \alpha_3 \\
       \alpha_5
 \end{pmatrix}
= R
\label{eq:lineareq}
\ee
where 
\be
R^T = [a^2 \alpha_a, c^2 \alpha_c, ac (\alpha_a + \alpha_c )\cos\beta - ac \beta \alpha_{\beta} \sin\beta ]
\ee
An inversion of \Eq~\ref{eq:lineareq} leads to 
\be
 \begin{pmatrix}
       \alpha_1 \\
       \alpha_3 \\
       \alpha_5
 \end{pmatrix}
= \frac{1}{( a c \sin\beta)^2} 
  \begin{pmatrix}
       c_3^2 &  a_3^2 &  -a_3 c_3\\
       -2 c_1 c_3 &  -2 a_1 a_3 &  a_1 c_3 + a_3 c_1 \\
       c_1^2 &  a_1^2  &  -a_1 c_1
  \end{pmatrix}
R
\label{eq:alphas}
\ee

Recall all quantities on the right hand side of \Eq~\ref{eq:alphas}
depend on $T$, we then have explicit (albeit complicated) expressions
for $\alpha_1(T)$, $\alpha_3(T)$, and $\alpha_5(T) $.  It may be
pointed out that these expressions for TEC tensor components are
somewhat different from that found in
Ref.~\onlinecite{Schlenker75v60}, presumably due to the fact that they
always orient the $c$ lattice parameter so that it is pointing in the
$z$ Cartesian direction.  Specifically they have $\alpha_2(T) =
\alpha_b(T)$ and $\alpha_3(T) = \alpha_c(T)$.  However, we do not
impose such condition but to allow the crystal to freely change its
shape under the effect of temperature while keeping the Cartesian
coordinates fixed.

\end{document}